\documentstyle[seceq,epsf,preprint]{ptptex}

\newcommand{\bra}[1]{\langle {#1} |}     
\newcommand{\ket}[1]{| {#1} \rangle}     
\newcommand{\bbra}[1]{\langle\!\langle {#1} |}     
\newcommand{\kket}[1]{| {#1} \rangle\!\rangle}     
\newcommand{\dkket}[1]{|\!| {#1} \rangle\!\rangle}     
\newcommand{\rket}[1]{| {#1} )}     
\newcommand{\drket}[1]{|\!| {#1} )}     
\newcommand{\drrket}[1]{|\!| {#1} )\!)}     
\newcommand{\dbra}[1]{\langle {#1} |\!|}     
\newcommand{\drbra}[1]{( {#1} |\!|}     
\newcommand{\dket}[1]{|\!| {#1} \rangle}     
\newcommand{\wtilde}[1]{\widetilde{#1}} 

\title{
On the Eigenvalue Problem of the $su(1,1)$-Algebra\\
and the Coupling Scheme of 
Two $su(1,1)$-Spins}

\author{
Seiya {\sc Nishiyama}$^{1}$, 
Constan\c{c}a {\sc Provid\^encia},$^{2}$\\
Jo\~ao da {\sc Provid\^encia}$^{2}$, Yasuhiko {\sc Tsue}$^{1}$ 
and Masatoshi {\sc Yamamura}$^{3}$
}

\inst{
$^{1}$Physics Division, Faculty of Science, Kochi University, Kochi 780-8520, 
Japan\\
$^{2}$Departamento de Fisica, Universidade de Coimbra, P-3000 Coimbra, 
Portugal\\
$^{3}$Faculty of Engineering, Kansai University, Suita 564-8680, Japan
}

\recdate{
\today
}

\abst{
After recapitulating the eigenvalue problem of the $su(1,1)$-algebra 
in the conventional form, the same problem is treated in an unconventional 
form, in which the eigenvalue is pure imaginary. 
Further, the coupling scheme of two $su(1,1)$-spins is discussed 
in the framework of two possibilities, in which certain new aspects appear. 
Finally, the coupling scheme developed in this paper is applied to a concrete 
example, which will serve boson realization of the $so(4)$- and the 
$so(3,1)$-algebra presented in the next paper. 
}

\begin{document}

\maketitle

\section{Introduction}

Boson realization of the Lie algebra has played a central role in theoretical 
studies of nuclear many-body systems. We can find its example 
in classical review article by Klein and Marshalek.\cite{1} 
Many of these studies are restricted to compact algebras such as 
the $su(2)$- and the $su(3)$-algebra.\cite{2,3} 
In the case of non-compact algebras, the $su(1,1)$-algebra 
is in quite interesting position, because this algebra enables us to 
describe the damped and the amplified harmonic oscillation in 
thermal circumstance as a conservative form.\cite{4,5} 
However, the $su(1,1)$-algebra seems for us to be fundamental not only 
for applying the above-mentioned problem but also for investigating the 
boson realization of various Lie algebras.\cite{6} 
In the next paper, we intend to report a possible boson realization of 
the $so(4)$- and the $so(3,1)$-algebra. 
In this case, also the $su(1,1)$-algebra plays a central role. 
In this boson realization, we encounter two points which are clarified 
for the $su(1,1)$-algebra. One is the eigenvalue problem in an 
unconventional form, which gives us a pure imaginary eigenvalue. 
Second is related to the coupling of two and three $su(1,1)$-spins.

With the aim of serving the boson realization of the $so(4)$- and the 
$so(3,1)$-algebra, the above-mentioned two points are discussed in this paper. 
The first point was already investigated in Ref. \citen{4} in rather 
formal form. Then, we borrow its basic parts. 
Concerning the second, we already developed the idea of the coupling scheme in 
the Holstein-Primakoff boson representation.\cite{7} 
In the present paper, we give a form which does not depend on the concrete 
representation, together with some aspects. 

In the next section, the eigenvalue problem in a conventional form is 
summarized. In \S 3, the eigenvalue problem is discussed in an unconventional 
form, in which pure imaginary eigenvalue appears. 
Section 4 is devoted to presenting the coupling of two $su(1,1)$-spins 
in two schemes, one of which gives us new aspects. 
In \S 5, a concrete example is treated. This example will play a central 
role in the next paper. 
Finally, a few comments are mentioned.

\section{Preliminary argument}

First of all, let us recapitulate basic part of the eigenvalue 
problem of a single $su(1,1)$-spin system. 
Hereafter, we call a system obeying the $su(1,1)$-algebra the $su(1,1)$-spin 
system. 
The $su(1,1)$-algebra is composed of three operators which we denote as 
${\hat T}_{\pm,0}$. 
They obey the relations 
\begin{eqnarray}
& &{\hat T}_0^*={\hat T}_0 \ , \qquad
{\hat T}_{\pm}^* ={\hat T}_{\mp} \ , 
\label{2-1}\\
& &[\ {\hat T}_+ \ ,\ {\hat T}_- \ ]=-2{\hat T}_0 \ , \qquad
[\ {\hat T}_0 \ ,\ {\hat T}_\pm \ ]=\pm{\hat T}_\pm \ . 
\label{2-2}
\end{eqnarray}
The Casimir operator, which we denote ${\hat{\mib T}}^2$, and 
its property are given by 
\begin{eqnarray}
& &{\hat{\mib T}}^2={\hat T}_0^2-(1/2)\cdot
({\hat T}_-{\hat T}_+ + {\hat T}_+{\hat T}_-)
={\hat T}_0({\hat T}_0\mp 1)-{\hat T}_{\pm}{\hat T}_{\mp} \ , 
\label{2-3}\\
& &[\ {\hat T}_{\pm,0} \ ,\ {\hat {\mib T}}^2 \ ]=0 \ .
\label{2-4}
\end{eqnarray}
A typical example of ${\hat T}_{\pm,0}$ is the following : 
\begin{equation}\label{2-5}
{\hat T}_+={\hat c}^*{\hat d}^* \ , \quad
{\hat T}_-={\hat d}{\hat c} \ , \quad
{\hat T}_0=1/2+(1/2)\cdot({\hat c}^*{\hat c}+{\hat d}^*{\hat d}) \ . 
\end{equation}
This expression was used for the description of the damped 
and the amplified harmonic oscillation.\cite{4,5}

We start in the assumption of the existence of the state $\ket{t(q)}$ 
which follows the relation 
\begin{eqnarray}\label{2-6}
& &{\hat T}_-\ket{t(q)}=0 \ , \qquad 
{\hat T}_0\ket{t(q)}=t \ket{t(q)} \ , \nonumber\\
& &\langle t(q) | t'(q')\rangle = \delta_{tt'}\delta_{qq'} \ .
\end{eqnarray}
Here, the symbol $(q)$ denotes a set of quantum numbers additional 
to $t$. For the system shown in the relation (\ref{2-5}) 
except the vacuum $\ket{0}$, the states which satisfy the relation (\ref{2-6}) 
are classified into two types, that is, 
$(\sqrt{(2t-1)!})^{-1}({\hat c}^*)^{2t-1}\ket{0}$ and 
$(\sqrt{(2t-1)!})^{-1}({\hat d}^*)^{2t-1}\ket{0}$ $(t=1, 3/2, 2,\cdots)$. 
The case $t=1/2$ corresponds to $\ket{0}$. This implies 
that, in order to discriminate between both types, the extra quantum number 
is necessary. In the case where we regard the sum of several 
$su(1,1)$-spin systems as a single system, it is natural that the 
extra quantum numbers are necessary. The relation (\ref{2-6}) leads us to 
\begin{equation}\label{2-7}
{\hat {\mib T}}^2\ket{t(q)}=t(t-1)\ket{t(q)} \ . 
\end{equation}
Since $\bra{t(q)}{\hat T}_-{\hat T}_+\ket{t(q)}$ is positive-definite and also 
$\bra{t(q)}{\hat T}_+{\hat T}_-\ket{t(q)}$ vanishes, $t$ should be positive. 
It is clear from the relation 
\begin{eqnarray}\label{2-8}
t&=&\bra{t(q)}{\hat T}_0\ket{t(q)}=(1/2)\bra{t(q)}[\ {\hat T}_- \ , \ 
{\hat T}_+\ ]\ket{t(q)} \nonumber\\
&=&(1/2)\bra{t(q)}{\hat T}_-{\hat T}_+\ket{t(q)} > 0 \ . 
\end{eqnarray}
The case $t=0$ is not interesting, because, in addition to 
${\hat T}_-\ket{t(q)}=0$, we have ${\hat T}_+\ket{t(q)}=0$ and there does not
exist any state connecting with $\ket{t=0(q)}$ through 
${\hat T}_{\pm}$.
Further, we note the following fact : Two positive values $t'$ and $t$ 
with $t'+t=1$ give us the relation $t'(t'-1)=t(t-1)$. 
This implies that, except the case $t'=t=1/2$, two different eigenvalues 
of ${\hat T}_0$ for $\ket{t(q)}$ and $\ket{t'=1-t(q)}$ give the same 
eigenvalue for ${\hat {\mib T}}^2$ in the region $0<t', t<1$. 
In the general framework, it may be impossible to fix the values of $t$, 
for example, in the case of the $su(2)$-algebra where $s$ is fixed 
$0,\ 1/2,\ 1,\ 3/2,\cdots$. 
It depends on the concrete cases, for example, $t=1/2,\ 1,\ 3/2,\cdots$ for 
the system (\ref{2-5}).

By operating ${\hat T}_+$ successively for $(t_0-t)$ times, we 
can define the following state $\ket{q;t,t_0}$ : 
\begin{eqnarray}\label{2-9}
& &\ket{q;t,t_0}=\sqrt{\frac{\Gamma(2t)}{(t_0-t)!\Gamma(t_0+t)}}
({\hat T}_+)^{t_0-t}\ket{t(q)}\ , \nonumber\\
& &\bra{q;t,t_0}=\sqrt{\frac{\Gamma(2t)}{(t_0-t)!\Gamma(t_0+t)}}
\bra{t(q)}({\hat T}_-)^{t_0-t}\ , \nonumber\\
& &\bra{q;t,t_0}q;t,t_0\rangle =1 \ . 
\end{eqnarray}
Here, $\Gamma(z)$ denotes the gamma function $(z=2t,\ t_0+t)$. 
The ket state $\ket{q;t,t_0}$ is an eigenstate of ${\hat {\mib T}}^2$ and 
${\hat T}_0$ and forms the orthogonal set : 
\begin{subequations}\label{2-10}
\begin{eqnarray}
& &{\hat {\mib T}}^2 \ket{q;t,t_0} = t(t-1)\ket{q;t,t_0} \ , 
\label{2-10a}\\
& &{\hat {T}_0} \ket{q;t,t_0} = t_0\ket{q;t,t_0} \ , 
\label{2-10b}
\end{eqnarray}
\end{subequations}
\begin{equation}\label{2-11}
\qquad
t_0=t, \ t+1, \ t+2, \cdots \ . 
\end{equation}
Of course, the bra state $\bra{q;t,t_0}$ also satisfies 
\begin{subequations}\label{2-12}
\begin{eqnarray}
& &\bra{q;t,t_0}{\hat {\mib T}}^2  = t(t-1)\bra{q;t,t_0} \ , 
\label{2-12a}\\
& &\bra{q;t,t_0}{\hat T}_0= t_0\bra{q;t,t_0} \ . 
\label{2-12b}
\end{eqnarray}
\end{subequations}

In the present system, we introduce the following operator : 
\begin{equation}\label{2-13}
{\hat Q}_+={\hat T}_+({\hat T}_0+{\hat T}+\varepsilon)^{-1} \ , \qquad
{\hat Q}_-=({\hat T}_0+{\hat T}+\varepsilon)^{-1}{\hat T}_- \ . 
\end{equation}
Here, $\varepsilon$ denote infinitesimal parameter which avoids null 
denominator and ${\hat T}$ is an operator obeying 
\begin{equation}\label{2-14}
{\hat T}\ket{q;t,t_0}=t\ket{q;t,t_0} \ . 
\end{equation}
Formally, it can be expressed in the form 
\begin{equation}\label{2-15}
{\hat T}=-{\hat {\mib T}}^2\left(1+\sum_{n=1}^{\infty}
\frac{(-)^n 2^n(2n-1)!!}{(n+1)!}\left({\hat {\mib T}}^2\right)^n \right) \ .
\end{equation}
Important property of ${\hat Q}_+$ is as follows : 
\begin{eqnarray}\label{2-16}
& &[\ {\hat T}_- \ , \ {\hat Q}_+\ ]=1 \ , \quad
[\ {\hat T}_0 \ , \ {\hat Q}_+\ ]={\hat Q}_+ \ , \quad
[\ {\hat T}_+ \ , \ {\hat Q}_+\ ]=({\hat Q}_+)^2 \ , \nonumber\\
& &\left[
\ \sqrt{{\hat T}_0+{\hat T}}\ {\hat Q}_- \ , \ {\hat Q}_+\sqrt{{\hat T}_0
+{\hat T}}\ \right]=1 \ . 
\end{eqnarray}
In the next problem and in the discussion on the addition of two 
$su(1,1)$-spins, ${\hat Q}_+$ will play a central role.

In addition to the eigenvalue equation (\ref{2-10b}), we can set up 
an eigenvalue problem given in the following relation : 
\begin{equation}\label{2-10c}
{\hat T}^0 \ket{q;t,t^0}=t^0\ket{q;t,t^0} \ . 
\end{equation}
Here, ${\hat T}^0$ is defined as 
\begin{equation}\label{2-17}
{\hat T}^0=(1/2)({\hat T}_+ + {\hat T}_-) \ . 
\end{equation}
Of course, $t^0$ is regarded as real. 
The properties of ${\hat T}^0$, together with ${\hat T}^{\pm}$ defined in 
the next section, will be presented in the next section for the problem 
of pure imaginary eigenvalue of ${\hat T}^0$. 
With the aim of obtaining the state $\ket{q;t,t^0}$, first, we investigate 
the following eigenvalue equation : 
\begin{equation}\label{2-19}
{\hat T}_-\ket{t,t^0(q)}=2t^0\ket{t,t^0(q)} \ . 
\end{equation}
With the use of the relation (\ref{2-16}), the normalized state 
$\ket{t,t^0(q)}$ is obtained in the form 
\begin{subequations}\label{2-20}
\begin{eqnarray}
& &\ket{t,t^0(q)}=\left(\sqrt{N(t,t^0)}\right)^{-1}
\exp\left(2t^0{\hat Q}_+\right)\ket{t(q)} \ , 
\label{2-20a}\\
& &N(t,t^0)=\sum_{n=0}^{\infty}\frac{(2t^0)^{2n}}{n!}\cdot
\frac{\Gamma(2t)}{\Gamma(2t+n)} \ . 
\label{2-20b}
\end{eqnarray}
\end{subequations}
Clearly, $t^0$ is arbitrary real number and $N(t,t^0)$ is convergent 
in the region $|t^0|<\infty$. 
Next, let the operator ${\hat \Omega}_+$, which obeys the 
following relation, exist : 
\begin{equation}\label{2-21}
{\hat T}_+{\hat \Omega}_+ + [{\hat T}_- \ , \ {\hat \Omega}_+]=0 \ .
\end{equation}
Then, we can verify easily that the state satisfying the eigenvalue 
equation (\ref{2-10c}), $\ket{q;t,t^0}$, is expressed in the form 
\begin{equation}\label{2-22}
\ket{q;t,t^0}={\hat \Omega}_+\ket{t,t^0(q)} \ . 
\end{equation}
The above means that our problem is reduced to find the explicit form of 
${\hat \Omega}_+$.

Let ${\hat \Omega}_+$ express in the form 
\begin{eqnarray}
& &{\hat \Omega}_+=\sum_{n=0}^{\infty}\frac{(-)^n}{n!}({\hat Q}_+)^{2n}
f_n({\hat T},{\hat T}_0-{\hat T}) \ , 
\label{2-23}\\
& &f_0({\hat T},{\hat T}_0-{\hat T})=1 \ . 
\label{2-24}
\end{eqnarray}
Substituting the expression (\ref{2-23}) into the relation (\ref{2-21}) 
and using the definition of ${\hat Q}_+$ shown in the relation (\ref{2-16}) 
and the relations (\ref{2-2})$\sim$(\ref{2-4}), we have 
\begin{eqnarray}\label{2-25}
& &({\hat T}_0-{\hat T}+2n)f_n({\hat T},{\hat T}_0-{\hat T})-({\hat T}_0
-{\hat T})f_n({\hat T},{\hat T}_0-{\hat T}-1) \nonumber\\
& &\qquad\qquad\qquad
-n({\hat T}_0-{\hat T}+2({\hat T}+n-1))f_{n-1}({\hat T},{\hat T}_0-{\hat T})
=0 \ . \qquad
(n=0, 1,2,\cdots) \nonumber\\
& &
\end{eqnarray}
By operating the state given in the relation (\ref{2-9}), $\ket{q;t,t_0}$, 
on the relation (\ref{2-25}), the following relation is obtained : 
\begin{eqnarray}
& &(m+2n)f_n(t,m)-mf_n(t,m-1)-n(m+2(t+n-1))f_{n-1}(t,m)=0 \ , 
\nonumber\\
& &m=t_0-t \ , \qquad (m=0,1,2,\cdots) 
\label{2-26}\\
& &f_0(t,m)=1\ . 
\label{2-27}
\end{eqnarray}
The form (\ref{2-26}) is a double recursion relation for $n$ and $m$ and 
successively from $n=m=0$, we obtain $f_n(t,m)$. 
Then, replacing $t$ and $m$ with ${\hat T}$ and ${\hat T}_0-{\hat T}$, 
respectively, $f_n({\hat T},{\hat T}_0-{\hat T})$ is derived. 
In Appendix, we will show the procedure for determining $f_n(t,m)$ and 
some lower cases are given. 
The above is a possible idea to solve the eigenvalue equation (\ref{2-7}). 
The state $\ket{q;t,t^0}$ shown in the relation (\ref{2-22}) may be 
not normalizable.

\section{An unconventional form of the eigenvalue problem}

In \S 2, we recapitulated the eigenvalue problem of the $su(1,1)$-spin 
system, in which ${\hat {\mib T}}^2$ and ${\hat T}_0$ are diagonalized. 
Further, in associating this eigenvalue problem, we sketched the case 
of $({\hat {\mib T}}^2 , {\hat T}_0)$. 
Concerning this problem, the following eigenvalue equation can be also 
set up : 
\begin{equation}\label{3-1}
{\hat T}^0\kket{q;t,t^0}=it^0\kket{q;t,t^0} \ . \qquad (i=\sqrt{-1})
\end{equation}
Here, ${\hat T}^0$ is defined in the relation (\ref{2-17}). 
The operators ${\hat T}^{\pm}$, which are used later, are 
defined as 
\begin{equation}\label{3-2}
{\hat T}^0=(1/2)\cdot ({\hat T}_++{\hat T}_-) \ , \qquad
{\hat T}^{\pm}=(1/2i)\cdot ({\hat T}_+-{\hat T}_-)\mp{\hat T}_0 \ .
\end{equation}
They obey the relations 
\begin{eqnarray}
& &{\hat T}^{0*}={\hat T}^0 \ , \qquad 
{\hat T}^{\pm *}={\hat T}^{\pm} \ , 
\label{3-3}\\
& &[\ {\hat T}^+\ , \ {\hat T}^-\ ]=2i{\hat T}^0 \ , \qquad
[\ {\hat T}^0\ , \ {\hat T}^\pm\ ]=\pm i{\hat T}^\pm \ . 
\label{3-4}
\end{eqnarray}
The Casimir operator ${\hat {\mib T}}^2$ and its property are given by 
\begin{eqnarray}
& &{\hat{\mib T}}^2=({\hat T}^0)^2-(1/2)\cdot
({\hat T}^-{\hat T}^+ + {\hat T}^+{\hat T}^-)
={\hat T}^0({\hat T}^0\mp i)-{\hat T}^{\pm}{\hat T}^{\mp} \ , 
\label{3-5}\\
& &[\ {\hat T}^{\pm,0} \ ,\ {\hat {\mib T}}^2 \ ]=0 \ .
\label{3-6}
\end{eqnarray}
The relations (\ref{3-3})$\sim$(\ref{3-6}) should be compared with the 
relations (\ref{2-1})$\sim$(\ref{2-4}). 
The state $\kket{q;t,t^0}$ satisfies 
\begin{equation}\label{3-7}
{\hat {\mib T}}^2\kket{q;t,t^0}=t(t-1)\kket{q;t,t^0}\ . 
\end{equation}
In the subsequent paper, we will encounter Eq.(\ref{3-1}) at some occasions. 
Since $t^0$ is proved to be real, Eq.(\ref{3-1}) does not claim the 
eigenvalue equation in the conventional sense.

Various properties of the relation (\ref{3-1}) can be found in Ref.\citen{4}, 
in which, instead of ${\hat T}^0$, the operator 
$(1/2i)\cdot({\hat T}_+-{\hat T}_-)$ was adopted. For the completeness of 
the paper, we borrow the results of Ref.\citen{4}. 
The $su(1,1)$-algebra in the representation $({\hat T}^{\pm,0})$ plays a 
central role in the case of describing ``damped and amplified harmonic 
oscillator" in thermal circumstance.\cite{4,5} 

The relation (\ref{3-7}) tells us that $t$ is a good quantum number and 
the state $\kket{q;t,t^0}$ satisfying the relation (\ref{3-1}) may be 
expanded in terms of the orthogonal set $\{\ket{q;t,t_0}\}$. 
First, we search the solutions of Eq.(\ref{3-1}) which obey the condition 
\begin{equation}\label{3-8-0}
{\hat T}^{\mp}\kket{q;t,t^0}=0 \ .
\end{equation}
Successive use of the relation (\ref{3-4}) gives us the solutions of 
Eq.(\ref{3-1}), which we denote $\kket{\pm t(q)}$, in the following form : 
\begin{subequations}
\begin{eqnarray}
& &\kket{\pm t(q)}=\left(\sqrt{N_t}\right)^{-1}\exp (\pm i{\hat T}_+)
\ket{t(q)} \ , \qquad t^0=\pm t \ , 
\label{3-8}\\
& &{\hat T}^0\kket{\pm t(q)}=\pm it\kket{\pm t(q)} \ , \qquad
{\hat T}^{\mp}\kket{\pm t(q)}=0 \ . 
\label{3-8a}
\end{eqnarray}
\end{subequations}
The quantity $N_t$ denotes the normalization constant, which will be 
discussed later. 
Further, we define the state 
\begin{equation}\label{3-9}
\kket{q;t,\pm t^0}=\left(\sqrt{N_{t,t^0}}\right)^{-1}
({\hat T}^{\pm})^{t^0-t}\kket{\pm t(q)} \ . \qquad (t^0 \geq t)
\end{equation}
Here, $N_{t,t^0}$ denotes also the normalization constant, 
which will be discussed later. The relation (\ref{3-4}) presents us 
the following relation : 
\begin{equation}\label{3-10}
{\hat T}^0\kket{q;t,\pm t^0}=\pm it^0\kket{q;t,\pm t^0} \ . 
\end{equation}
The above indicates that the state (\ref{3-9}) is identical with 
the solution of Eq.(\ref{3-1}).

Our next task is to clarify the orthogonality of the set 
$\{\kket{q;t,t^0}\}$. Since $t^0$ in Eq.(\ref{3-1}) is real, i.e., $it^0$ is 
pure imaginary, it is impossible to define the orthogonality 
in the conventional manner. 
In Ref.\citen{4}, on the basis of the time-reversal conjugate, the problem was 
discussed and in this paper, we translate the basic part of Ref.\citen{4} 
in a plain form. We define the bra state conjugate to the ket state 
$\kket{q;t,t^0}$ in the form 
\begin{eqnarray}
& &\bbra{q;t,\pm t^0}=\left(\sqrt{N_{t,t^0}}\right)^{-1}
\bbra{\pm t(q)}({\hat T}^{\mp})^{t^0-t} \ , 
\label{3-11}\\
& &\bbra{\pm t(q)}=\left(\sqrt{N_t}\right)^{-1}\bra{t(q)}\exp(\pm i{\hat T}_-)
\ . 
\label{3-12}
\end{eqnarray}
The state $\bbra{q;t,\pm t^0}$ satisfies 
\begin{subequations}
\begin{equation}\label{3-13}
\bbra{q;t,\pm t^0}{\hat T}^0 = \pm it^0\bbra{q;t,\pm t} \ . 
\end{equation}
Further, the state $\bbra{\pm t(q)}$ obeys 
\begin{equation}\label{3-13a}
\bbra{\pm t(q)}{\hat T}^0 = \pm it \bbra{\pm t(q)} \ , \qquad
\bbra{\pm t(q)}{\hat T}^{\pm}=0 \ . 
\end{equation}
\end{subequations}
The definitions (\ref{3-9}) and (\ref{3-11}) for the ket and the bra states, 
$\kket{q;t,\pm t^0}$ and $\bbra{q;t,\pm t^{0'}}$, gives us the relation 
\begin{equation}\label{3-14}
\bbra{q;t,\pm t^{0'}} q;t,\pm t^0 \rangle\!\rangle =0 \ . 
\qquad (t^{0'}\neq t^0)
\end{equation}
However, the following relation should be noted : 
\begin{equation}\label{3-15}
\bbra{q;t,\pm t^{0'}} q;t,\mp t^0 \rangle\!\rangle \neq 0 \ . 
\qquad (t^{0'}\neq t^0)
\end{equation}
For the other quantum numbers, it may be trivial. 
Concerning the normalization, we set up the relations 
\begin{eqnarray}
& &N_{t,t^0}\bbra{q;t,\pm t^{0}} q;t,\pm t^0 \rangle\!\rangle \nonumber\\
& &\qquad =\bbra{\pm t(q)}({\hat T}^{\mp})^{t^0-t}({\hat T}^{\pm})^{t^0-t}
\kket{\pm t(q)} \ , 
\label{3-16}\\
& &N_{t}\bbra{\pm t(q)} \pm t(q) \rangle\!\rangle \nonumber\\
& &\qquad =\bra{t(q)}e^{\pm i{\hat T}_-}e^{\pm i{\hat T}_+}
\ket{t(q)} \ . 
\label{3-17} 
\end{eqnarray}
Successive use of the relation (\ref{3-4}) gives us 
\begin{eqnarray}
& &N_{t,t^0}\bbra{q;t,\pm t^{0}} q;t,\pm t^0 \rangle\!\rangle \nonumber\\
& &\qquad =\frac{(t^0-t)!\Gamma(t^0+t)}{\Gamma(2t)}
\bbra{\pm t(q)}{\pm t(q)}\rangle\!\rangle \ , 
\label{3-18}\\
& &N_{t}\bbra{\pm t(q)} \pm t(q) \rangle\!\rangle 
=\sum_{n=0}^{\infty}\frac{(-)^n\Gamma(n+2t)}{n!\Gamma(2t)} \ . 
\label{3-19} 
\end{eqnarray}
We note the right-hand side of the relation (\ref{3-19}). 
This is a kind of the hypergeometric series and it is not absolutely 
convergent. However, we have 
\begin{equation}\label{3-20}
\sum_{n=0}^{\infty}\frac{(-)^n \Gamma(n+2t)}{n!\Gamma(2t)} x^n=(1+x)^{-2t}
 \ . \qquad (|x|<1)
\end{equation}
Then, it may be conjectured to replace the right-hand side of the relation 
(\ref{3-19}) with $2^{-2t}$ and through this process, we set up 
\begin{eqnarray}
& &N_t = 2^{-2t} \ , \qquad {\rm i.e.,}\qquad 
\bbra{\pm t(q)}\pm t(q)\rangle\!\rangle=1 \ , 
\label{3-21}\\
& &N_{t,t^0}=\frac{(t^0-t)!\Gamma(t^0+t)}{\Gamma(2t)} \ , 
\quad {\rm i.e.,}\quad 
\bbra{q;t,\pm t^0}q;t,\pm t^0 \rangle\!\rangle=1 \ .
\end{eqnarray}
In the sense mentioned above, the state $\kket{q;t,\pm t^0}$ is normalized.

The state $\kket{q;t,\pm it^0}$ may be an eigenstate of the operators 
${\hat Q}$, ${\hat T}$ and ${\hat T}^0$ with the eigenvalues $q$, $t$ and 
$\pm it^0$, which gives us the eigenstate of the Hamiltonian 
${\hat H}=E({\hat Q}; {\hat T},{\hat T}^0)$ as a function of 
${\hat Q}$, ${\hat T}$ and ${\hat T}^0$ : 
\begin{equation}\label{2-24x}
{\hat H}\kket{q;t,\pm it^0}=E(q;t,\pm it^0)\kket{q;t,\pm it^0} \ .
\end{equation}
Here, it should be noted that the eigenvalue $E(q;t,\pm it^0)$ is not always 
real. 
In the case where we treat the time-evolution of the system obeying the 
Hamiltonian ${\hat H}=E({\hat Q};{\hat T},{\hat T}^0)$, the solution 
of the following Schr\"odinger equation should be investigated : 
\begin{equation}\label{2-25x}
i\partial_{\tau}\kket{\psi(\tau)}={\hat H}\kket{\psi(\tau)} \ . \qquad
(\tau;\ {\rm the\ variable\ of\ time})
\end{equation}
As a possible solution of Eq.(\ref{2-25x}), we can adopt 
\begin{equation}\label{2-26x}
\kket{q;t,\pm it^0(\tau)}=\kket{q;t,\pm it^0}e^{-iE(q;t,\pm it^0)\tau} \ .
\end{equation}
Further, we require the bra-state of $\kket{q;t,\pm it^0(\tau)}$ 
in the form 
\begin{equation}\label{2-27x}
\bbra{q;t,\pm it^0(\tau)}=\bbra{q;t,\pm it^0}e^{iE(q;t,\pm it^0)^* \tau} \ .
\end{equation}
In the form (\ref{2-27x}), $E^*$ denotes complex conjugate of $E$. 
The bra- and the ket-state give us 
\begin{equation}\label{2-28x}
\bbra{q;t,\pm it^0(\tau)}q;t,\pm it^0(\tau)\rangle\!\rangle
=e^{2({\rm Im}E(q;t,\pm it^0))\tau} \ . 
\end{equation}
Here, ${\rm Im}\ E$ denotes the imaginary part of $E$. 
General solution of Eq.(\ref{2-25x}) is expressed as 
\begin{equation}\label{2-29x}
\kket{\psi(\tau)}=\sum_{q,t,t^0}C_{q;t,t^0}\kket{q;t,\pm it^0}
e^{-iE(q;t,\pm it^0)\tau} \ . 
\end{equation}
Here, the coefficient $C_{q;t,t^0}$ is determined by the initial condition. 
We require the condition 
\begin{equation}\label{2-30x}
\sum_{q,t,t^0}|C_{q;t,t^0}|^2=1 \ . 
\end{equation}
The bra-state is defined as 
\begin{equation}\label{2-31x}
\bbra{\psi(\tau)}=\sum_{q,t,t^0}C_{q;t,t^0}^*\bbra{q;t,\pm it^0}
e^{iE(q;t,\pm it^0)^*\tau} \ . 
\end{equation}
Then, the norm $\bbra{\psi(\tau)}\psi(\tau)\rangle\!\rangle$ is given 
in the form 
\begin{equation}\label{2-30y}
\bbra{\psi(\tau)}\psi(\tau)\rangle\!\rangle
=\sum_{q,t,t^0}|C_{q;t,t^0}|^2 e^{2({\rm Im}E(q;t,\pm it^0))\tau} \ .
\end{equation}
Here, ${\rm Im}E(q;t,\pm it^0)$ denotes imaginary part of $E(q;t,\pm it^0)$. 
Of course, the condition (\ref{2-30y}) gives us 
\begin{equation}\label{2-31y}
\bbra{\psi(0)}\psi(0)\rangle\!\rangle=1 \ . 
\end{equation}
The relation (\ref{2-30y}) suggests us that our present idea has a 
possibility of describing the damping and amplifying of motion 
of the system under investigation. 
If the norm (\ref{2-30y}) is decreasing for $\tau$, the system is 
under the damping and if increasing, the amplifying. 
In the forth coming paper, we will discuss this problem.


\section{Coupling of two $su(1,1)$-spins}

Under the preparation discussed in \S\S 2 and 3, we investigate the 
addition of two $su(1,1)$-spins. The present authors gave an idea 
for the case of the Holstein-Primakoff boson representation\cite{7} 
and the present treatment is, in some sense, independent of the 
representation and contains some new aspects. 
We denote the operators governing two $su(1,1)$-algebras as 
${\hat X}_{\pm,0}$ and ${\hat Y}_{\pm,0}$. 
The sum of two $su(1,1)$-algebras is defined in the following two forms :
\begin{subequations}\label{4-1}
\begin{eqnarray}
& &{\hat T}_{\pm,0}={\hat X}_{\pm,0}+{\hat Y}_{\pm,0} \ , 
\label{4-1a}\\
& &{\wtilde T}_{\pm,0}={\hat X}_{\pm,0}-{\hat Y}_{\mp,0} \ . 
\label{4-1b}
\end{eqnarray}
\end{subequations}
We call the forms (\ref{4-1a}) and (\ref{4-1b})  the $a$- and the $b$-type, 
respectively. It is easily verified that they obey 
\begin{subequations}\label{4-2}
\begin{eqnarray}
& & [\ {\hat T}_+\ , \ {\hat T}_-\ ]= -2{\hat T}_0 \ , \qquad
[\ {\hat T}_0\ , \ {\hat T}_\pm\ ]= \pm{\hat T}_\pm\ , 
\label{4-2a}\\
& & [\ {\wtilde T}_+\ , \ {\wtilde T}_-\ ]= -2{\wtilde T}_0 \ , \qquad
[\ {\wtilde T}_0\ , \ {\wtilde T}_\pm\ ]= \pm{\wtilde T}_\pm \ . 
\label{4-2b}
\end{eqnarray}
\end{subequations}
In the form analogous to the relation (\ref{2-6}), we set up 
\begin{subequations}\label{4-3}
\begin{eqnarray}
& & {\hat X}_-\dket{x(u)}=0 \ , \qquad {\hat X}_0\dket{x(u)}=x\dket{x(u)} \ , 
\quad (x>0) \nonumber\\
& &\dbra{x(u)}x'(u')\rangle =\delta_{xx'}\delta_{uu'} \ , 
\label{4-3a}\\
& & {\hat Y}_-\drket{y(v)}=0 \ , \qquad {\hat Y}_0\drket{y(v)}=y\drket{y(v)} 
\ , 
\quad (y>0) \nonumber\\
& &\drbra{y(v)}y'(v')) =\delta_{yy'}\delta_{vv'} \ . 
\label{4-3b}
\end{eqnarray}
\end{subequations}
Of course, $u$ and $v$ denote the extra quantum numbers. 
With the use of the two states $\dket{x(u)}$ and $\drket{y(v)}$, we define 
the states for the $a$- and $b$-type as follows : 
\begin{subequations}\label{4-4}
\begin{eqnarray}
& & \ket{x(u),y(v)}=\dket{x(u)}\otimes \drket{y(v)} \ , 
\label{4-4a}\\
& & \rket{x(u),y(v)}=\left(\sqrt{N_{xy}}\right)^{-1}\exp\left[
{\hat X}_+({\hat X}_0+{\hat X}+\varepsilon)^{-1}{\hat Y}_+\right] \nonumber\\
& &\qquad\qquad\qquad\times \dket{x(u)}\otimes \drket{y(v)} \ . 
\label{4-4b}
\end{eqnarray}
\end{subequations}
Here, $N_{xy}$ denotes the normalization constant which will be discussed 
later. 
The meaning of ${\hat X}$ and $\varepsilon$ may be understood from the 
discussion related to the relations (\ref{2-13})$\sim$(\ref{2-15}). 
The states $\ket{x(u),y(v)}$ and $\rket{x(u),y(v)}$ satisfy 
\begin{subequations}\label{4-5}
\begin{eqnarray}
& & {\hat T}_-\ket{x(u),y(v)}=0 \ , \qquad
{\hat T}_0\ket{x(u),y(v)}=(x+y)\ket{x(u),y(v)} \ , 
\label{4-5a}\\
& & {\wtilde T}_-\rket{x(u),y(v)}=0 \ , \qquad
{\wtilde T}_0\rket{x(u),y(v)}=(x-y)\rket{x(u),y(v)}\ . 
\label{4-5b}
\end{eqnarray}
\end{subequations}
The normalization constant $N_{xy}$ is calculated in the form 
\begin{equation}\label{4-6}
N_{xy}=1+\sum_{n=1}^{\infty}\frac{2y(2y+1)\cdots (2y+n-1)}
{2x(2x+1)\cdots (2x+n-1)}
=F(2y,1,2x;1) \ . 
\end{equation}
The symbol $F(2y,1,2x;1)$ denotes special case of Gauss' hypergeometric 
function $F(a,b,c;z)$ and it is given as 
\begin{equation}\label{4-7}
F(2y,1,2x;1)=\frac{\Gamma(2x)\Gamma(2x-2y-1)}{\Gamma(2x-1)\Gamma(2y-1)}
=\frac{2x-1}{2x-2y-1} \ . \quad 
(2x-2y-1>0)
\end{equation}
Then, combining with $x>0$ and $y>0$, the condition $N_{xy}>0$ leads us 
to the following condition, under which the state $\rket{x(u),y(v)}$ 
can be normalized :
\begin{equation}\label{4-8}
2x>2y+1>1 \ . \qquad {\rm (for\ the}\ b\hbox{\rm -type)}
\end{equation}

Under the above results, we obtain two states $\ket{x(u),y(v);t}$ and 
$\rket{x(u),y(v);t}$ which satisfy 
\begin{subequations}\label{4-9}
\begin{eqnarray}
& & {\hat T}_-\ket{x(u),y(v);t}=0 \ , \qquad
{\hat T}_0\ket{x(u),y(v);t}=t\ket{x(u),y(v);t} \ , 
\label{4-9a}\\
& & {\wtilde T}_-\rket{x(u),y(v);t}=0 \ , \qquad
{\wtilde T}_0\rket{x(u),y(v);t}=t\rket{x(u),y(v);t}\ . 
\label{4-9b}
\end{eqnarray}
\end{subequations}
The explicit forms are as follows : 
\begin{subequations}\label{4-10}
\begin{eqnarray}
& & \ket{x(u),y(v);t}=\left(\sqrt{N_{xy,t(a)}}\right)^{-1}
({\hat O}_+)^{t-(x+y)}\ket{x(u),y(v)} \ , 
\label{4-10a}\\
& & \rket{x(u),y(v);t}=\left(\sqrt{N_{xy,t(b)}}\right)^{-1}
({\wtilde O}_+)^{(x-y)-t}\rket{x(u),y(v)}\ . 
\label{4-10b}
\end{eqnarray}
\end{subequations}
\vspace{-0.6cm}
\begin{subequations}\label{4-11}
\begin{eqnarray}
& & {\hat O}_+={\hat X}_+({\hat X}_0+{\hat X}+\varepsilon)^{-1}-
{\hat Y}_+({\hat Y}_0+{\hat Y}+\varepsilon)^{-1} \ , 
\label{4-11a}\\
& & {\wtilde O}_+={\hat Y}_+ \ . 
\label{4-11b}
\end{eqnarray}
\end{subequations}
The operator ${\hat O}_+$ is closely related to ${\hat Q}_+$ defined 
in the relation (\ref{2-13}). The quantity $\varepsilon$ is an infinitesimal 
parameter and the definitions of ${\hat X}$ and ${\hat Y}$ may be 
understood from the relations (\ref{2-14}) and (\ref{2-15}). 
The operators ${\hat O}_+$ and ${\wtilde O}_+$ satisfy 
\begin{subequations}\label{4-12}
\begin{eqnarray}
& & [\ {\hat T}_- \ , \ {\hat O}_+\ ]=0 \ , \qquad 
[\ {\hat T}_0 \ , \ {\hat O}_+\ ]=+{\hat O}_+ \ ,
\label{4-12a}\\
& &  [\ {\wtilde T}_- \ , \ {\wtilde O}_+\ ]=0 \ , \qquad 
[\ {\wtilde T}_0 \ , \ {\wtilde O}_+\ ]=-{\wtilde O}_+\ . 
\label{4-12b}
\end{eqnarray}
\end{subequations}
With the use of the relation (\ref{4-12}), we obtain the relation (\ref{4-9}). 
The quantities $N_{xy,t(a)}$ and $N_{xy,t(b)}$ denote the 
normalization constants : 
\begin{subequations}\label{4-13}
\begin{eqnarray}
& & N_{xy,t(a)}=\sum_{r=0}^{t-(x+y)}\frac{(t-(x+y))!\Gamma(2x)\Gamma(2y)}
{r!\Gamma(t+x-y-r)\Gamma(2y+r)} \ , 
\label{4-13a}\\
& & N_{xy,t(b)}=\frac{((x-y)-t)!\Gamma(x+y-t)}
{\Gamma(2y)}\ . 
\label{4-13b}
\end{eqnarray}
\end{subequations}
The form (\ref{4-10}) tells us 
\begin{subequations}\label{4-14}
\begin{eqnarray}
& & t=x+y\ , \ x+y+1 \ , \ x+y+2 \ , \cdots \ , \quad ({\rm for\ the\ }a
\hbox{\rm -type)}
\label{4-14a}\\
& &  t=x-y\ , \ x-y-1 \ , \cdots \ , t_m \ . 
\quad ({\rm for\ the\ }b\hbox{\rm -type)}\ . 
\label{4-14b}
\end{eqnarray}
\end{subequations}
For the $b$-type, $x$ and $y$ obey the condition (\ref{4-8}) and 
$t_m$ denotes the minimum of $t$ ($0<t_m \leq 1$). 
The relations (\ref{4-14a}) and (\ref{4-14b}) serve us the coupling rule 
for the additions of two $su(1,1)$-spins. In Ref.\citen{7}, only the 
case (\ref{4-14a}) was discussed. Concerning ${\wtilde O}_+$, 
we give a  short comment. 
In the form analogous to ${\hat O}_+$, we can define formally 
${\wtilde O}_+'$ as ${\wtilde O}_+'={\hat X}_+({\hat X}_0+{\hat X}+
\varepsilon)^{-1}-({\hat Y}_0+{\hat Y}+\varepsilon)^{-1}{\hat Y}_-$. 
The operator ${\wtilde O}_+'$ obeys the same relation as that shown in 
the relation (\ref{4-12b}). However, we have 
${\wtilde O}_+'\rket{x(u),y(v)}=0$, and then, ${\wtilde O}_+'$ cannot play the 
same role as that of ${\wtilde O}_+$.

By operating ${\hat T}_+$ and ${\wtilde T}_+$ on the states 
$\ket{x(u),y(v);t}$ and $\rket{x(u),y(v);t}$ successively for $(t_0-t)$ times, 
respectively, we obtain 
\begin{subequations}\label{4-15}
\begin{eqnarray}
& & \ket{x(u),y(v);t,t_0}=\sqrt{\frac{\Gamma(2t)}{(t_0-t)!\Gamma(t_0+t)}}
({\hat T}_+)^{t_0-t}\ket{x(u),y(v);t} \ , 
\label{4-15a}\\
& & \rket{x(u),y(v);t,t_0}=\sqrt{\frac{\Gamma(2t)}{(t_0-t)!\Gamma(t_0+t)}}
({\wtilde T}_+)^{t_0-t}\rket{x(u),y(v);t}\ . 
\label{4-15b}
\end{eqnarray}
\end{subequations}
Of course, both states are normalized. We can prove that both states are 
eigenstates of ${\hat {\mib X}}^2$ and ${\hat {\mib Y}}^2$, and further, 
$({\hat {\mib T}}^2, {\hat T}_0)$ and 
$({\wtilde {\mib T}}^2 , {\wtilde T}_0)$, respectively, with the eigenvalues 
$x(x-1)$, $y(y-1)$, $t(t-1)$ and $t_0$. 
It should be noted that $t_0$ obeys the rule (\ref{2-12}) and $t$ is governed 
by the rule (\ref{4-14}). For the $a$-type, $x$ and $y$ are positive and for 
the $b$-type, $x$ and $y$ are also positive with the condition 
(\ref{4-8}). The above is the coupling scheme of two $su(1,1)$-algebras and 
the states $\ket{x(u),y(v);t,t_0}$ and $\rket{x(u),y(v);t,t_0}$ 
are monomial.

\section{An example}

In the next paper, we will describe possible boson realizations of the 
$so(4)$- and the $so(3,1)$-algebra, in which we contact the following 
$su(1,1)$-algebras : 
\begin{subequations}\label{5-1}
\begin{eqnarray}
& &{\hat T}_+
={\hat c}_+^*{\hat c}_-^*-(1/2){\hat c}_0^{*2}+(1/2){\hat d}_0^{*2} \ , 
\nonumber\\
& &{\hat T}_-
={\hat c}_-{\hat c}_+-(1/2){\hat c}_0^{2}+(1/2){\hat d}_0^{2} \ , 
\nonumber\\
& &{\hat T}_0
=1+(1/2)({\hat c}_+^*{\hat c}_+ + {\hat c}_0^*{\hat c}_0 + 
{\hat c}_-^*{\hat c}_-)+(1/2){\hat d}_0^{*}{\hat d}_0 \ , 
\label{5-1a}\\
& &{\wtilde T}_+
={\hat c}_+^*{\hat c}_-^*-(1/2){\hat c}_0^{*2}-(1/2){\hat d}_0^{2} \ , 
\nonumber\\
& &{\wtilde T}_-
={\hat c}_-{\hat c}_+ -(1/2){\hat c}_0^{2}-(1/2){\hat d}_0^{*2} \ , 
\nonumber\\
& &{\wtilde T}_0
=1/2+(1/2)({\hat c}_+^*{\hat c}_+ + {\hat c}_0^*{\hat c}_0 + 
{\hat c}_-^*{\hat c}_-)-(1/2){\hat d}_0^{*}{\hat d}_0 \ . 
\label{5-1b}
\end{eqnarray}
\end{subequations}
Here, $({\hat c}_{\pm,0}^* , {\hat c}_{\pm,0})$ and 
$({\hat d}_{0}^* , {\hat d}_{0})$ denote boson operators. 
As an example of the coupling scheme developed in \S 4, 
we treat the cases (\ref{5-1a}) and (\ref{5-1b}). The forms 
(\ref{5-1a}) and (\ref{5-1b}) can be decomposed into the forms 
(\ref{4-1a}) and (\ref{4-1b}), respectively : 
\begin{eqnarray}
& &{\hat X}_+={\hat c}_+^*{\hat c}_-^* - (1/2){\hat c}_0^{*2} \ , \qquad
{\hat X}_- ={\hat c}_-{\hat c}_+ -(1/2){\hat c}_0^2 \ , \nonumber\\
& &{\hat X}_0=3/4+(1/2)({\hat c}_+^*{\hat c}_+ + {\hat c}_0^*{\hat c}_0 + 
{\hat c}_-^*{\hat c}_-) \ , 
\label{5-2}\\
& &{\hat Y}_+=(1/2){\hat d}_0^{*2} \ , \qquad {\hat Y}_-=(1/2){\hat d}_0^2 \ , 
\qquad {\hat Y}_0=1/4+(1/2){\hat d}_0^*{\hat d}_0 \ . 
\label{5-3}
\end{eqnarray}
Further, ${\hat X}_{\pm,0}$ can be decomposed into the following form : 
\begin{eqnarray}
& &{\hat X}_{\pm,0}={\hat W}_{\pm,0}+{\hat Z}_{\pm,0} \ , 
\label{5-4}\\
& &{\hat W}_+={\hat c}_+^*{\hat c}_-^* \ , \qquad
{\hat W}_-={\hat c}_-{\hat c}_+ \ , \qquad 
{\hat W}_0=1/2+(1/2)({\hat c}_+^*{\hat c}_++{\hat c}_-^*{\hat c}_-) \ , 
\label{5-5}\\
& &{\hat Z}_+=-(1/2){\hat c}_0^{*2} \ , \qquad
{\hat Z}_-=-(1/2){\hat c}_0^2 \ , \qquad
{\hat Z}_0=1/4+(1/2){\hat c}_0^*{\hat c}_0 \ . 
\label{5-6}
\end{eqnarray}
The sets $({\hat W}_{\pm,0})$ and $({\hat Z}_{\pm,0})$ also obey the 
$su(1,1)$-algebras, respectively. 
It may be clear that the present system is the sum of three 
$su(1,1)$-algebras, $({\hat W}_{\pm,0})$, $({\hat Z}_{\pm,0})$ and 
$({\hat Y}_{\pm,0})$. Then, successively, the addition is performed.

Our first task is to search the state $\ket{x(u)}$ obeying the 
relation (\ref{4-3a}) under the sum (\ref{5-4}). 
It is completed by replacing $x$, $y$, $t$, $({\hat X}_{\pm,0}, {\hat X})$ 
and $({\hat Y}_{\pm,0} , {\hat Y})$ in the form (\ref{4-10a}) with 
$w$, $z$, $x$, $({\hat W}_{\pm,0} , {\hat W})$ and 
$({\hat Z}_{\pm,0} , {\hat Z})$, respectively. 
As for $u$ and $v$ in the form (\ref{4-10a}), we adopt 
$u=s=+, -$ and $v$ is not necessary. 
Later, we will show them. 
By denoting $\ket{w(s)z;x}$ as $\dket{x(w(s)z)}$, 
we obtain 
\begin{eqnarray}\label{5-7}
\dket{x(w(s)z)}=\left[{\hat W}_+({\hat W}_0+{\hat W}+\varepsilon)^{-1}
-{\hat Z}_+({\hat Z}_0+{\hat Z}+\varepsilon)^{-1}\right]^{x-(w+z)}
\ket{w(s),z} .\qquad
\end{eqnarray}
Here and hereafter, we omit the normalization constant for any state. 
The state $\ket{w(s),z}$ is obtained by the form (\ref{4-4a}) : 
\begin{eqnarray}
& &\ket{w(s),z}=\dkket{w(s)}\otimes \drrket{z} \ , 
\label{5-8}\\
& &{\hat W}_-\dkket{w(s)}=0 \ , \qquad 
{\hat W}_0\dkket{w(s)}=w\dkket{w(s)} \ , 
\label{5-9}\\
& &{\hat Z}_-\drrket{z}=0 \ , \qquad 
{\hat Z}_0\drrket{z}=z\drrket{z} \ . 
\label{5-10}
\end{eqnarray}
The states $\dkket{w(s)}$ and $\drrket{z}$ are concretely expressed as 
\begin{eqnarray}
& &\dkket{w(s)}=\cases{({\hat c}_+^*)^{2w-1}\dkket{0} \ , \quad
(s=+\ , w=1\ , \ 3/2 \ , \ 2 \ , \ 5/2 \ , \cdots) \cr
({\hat c}_-^*)^{2w-1}\dkket{0} \ , \quad
(s=-\ , w=1/2 \ , \ 1 \ , \ 3/2 \ , \ 2 \ ,  \cdots)
}
\label{5-11}\\
& &\drrket{z}=\cases{\drrket{0} \ , \quad\ \ 
(z=1/4) \cr
{\hat c}_0^*\drrket{0} \ . \quad
(z=3/4)
}
\label{5-12}
\end{eqnarray}
Here, $\dkket{0}$ and $\drrket{0}$ denote the vacuums for ${\hat c}_{\pm}$ 
and ${\hat c}_0$, respectively. 
The state $\dket{x(w(s)z)}$ satisfies the relation (\ref{4-3a}) : 
\begin{equation}\label{5-13}
{\hat X}_-\dket{x(w(s)z)}=0 \ , \qquad 
{\hat X}_0\dket{x(w(s)z)}=x\dket{x(w(s)z)} \ . 
\end{equation}
In the present case, the rule (\ref{4-14a}) gives us 
\begin{equation}\label{5-14}
x=w+z \ , \ w+z+1 \ , \ w+z+2 \ , \cdots \ .
\end{equation}
In the same form as the above, we can derive the state $\drket{y}$ satisfying 
the relation (\ref{4-3b}) : 
\begin{equation}\label{5-15}
\drket{y}=\cases{\drket{0} \ , \quad\ \ 
(y=1/4) \cr
{\hat d}_0^*\drket{0} \ . \quad
(y=3/4)
}
\end{equation}
Here, $\drket{0}$ denotes the vacuum for ${\hat d}_0$. 
In the present case, $v$ in the relation (\ref{4-3b}) is not necessary. 
The state $\drket{y}$ satisfies the relation (\ref{4-3b}) : 
\begin{equation}\label{5-16}
{\hat Y}_-\drket{y}=0 \ , \qquad {\hat Y}_0\drket{y}=y\drket{y} \ . 
\end{equation}
Thus, we obtain $\ket{x(w(s)z)y;t,t_0}$ and $\rket{x(w(s)z)y;t,t_0}$ 
in the following form : 
\begin{subequations}\label{5-17}
\begin{eqnarray}
& &\ket{x(w(s)z)y;t,t_0}=
({\hat T}_+)^{t_0-t}\cdot\left[{\hat X}_+({\hat X}_0+x+\varepsilon)^{-1}
-{\hat Y}_+({\hat Y}_0+y+\varepsilon)^{-1}\right]^{t-(x+y)} \nonumber\\
& &\qquad\qquad\qquad\qquad\qquad\qquad
\times\left[{\hat W}_+({\hat W}_0+w+\varepsilon)^{-1}
-{\hat Z}_+({\hat Z}_0+z+\varepsilon)^{-1}\right]^{x-(w+z)} \nonumber\\
& &\qquad\qquad\qquad\qquad\qquad\qquad
\times\ket{w(s),z,y} \ , 
\label{5-17a}\\
& &\rket{x(w(s)z)y;t,t_0}=
({\wtilde T}_+)^{t_0-t}\cdot({\hat Y}_+)^{(x-y)-t} 
\cdot \exp\left[{\hat X}_+({\hat X}_0+x+\varepsilon)^{-1}{\hat Y}_+\right]
\nonumber\\
& &\qquad\qquad\qquad\qquad\qquad\qquad
\times\left[{\hat W}_+({\hat W}_0+w+\varepsilon)^{-1}
-{\hat Z}_+({\hat Z}_0+z+\varepsilon)^{-1}\right]^{x-(w+z)} \nonumber\\
& &\qquad\qquad\qquad\qquad\qquad\qquad
\times\ket{w(s),z,y} \ , 
\label{5-17b}
\end{eqnarray}
\end{subequations}
\vspace{-0.5cm}
\begin{equation}\label{5-18}
\ket{w(s),z,y}=\dkket{w(s)}\otimes \drrket{z}\otimes \drket{y} \ . 
\qquad\qquad\qquad\qquad\qquad\qquad
\end{equation}
It may be natural that, following the given values of $s=+,\ -$, $z=1/4,\ 
3/4$ and $y=1/4, \ 3/4$, the values of $w$, $x$ and $t$ can be changed. 
In the next paper, we will describe the $so(4)$- and the 
$so(3,1)$-algebra, in which we encounter the case 
$(s=- , \ z=1/4, \ x=w+1/4)$ : 
\begin{subequations}\label{5-19}
\begin{eqnarray}
& &\ket{w+1/4(w(-)1/4)y;t,t_0}\nonumber\\
&=&
({\hat T}_+)^{t_0-t}\!\!\cdot\!
\left[{\hat X}_+({\hat X}_0+w+1/4+\varepsilon)^{-1}
\!\!-{\hat Y}_+({\hat Y}_0+y+\varepsilon)^{-1}\right]^{t-w-1/4-y} 
\ket{w(-),1/4,y} \ , \nonumber\\
& &\label{5-19a}\\
& &\rket{w+1/4(w(-)1/4)y;t,t_0}\nonumber\\
&=&
({\wtilde T}_+)^{t_0-t}\cdot({\hat Y}_+)^{w+1/4-y-t} 
\cdot \exp\left[{\hat X}_+({\hat X}_0+w+1/4+\varepsilon)^{-1}{\hat Y}_+\right]
\ket{w(-),1/4,y} \ , \nonumber\\
& &\label{5-19b}
\end{eqnarray}
\end{subequations}
Of course, the notations used in this paper are different from those used 
in the next paper.

\section{Discussion}
Finally, we investigate the structure of the states $\ket{x(w(s)z)y;t,t_0}$ 
and\break
$\rket{x(w(s)z)y;t,t_0}$ (hereafter, we call these two states simply 
the ``states (\ref{5-17})"). 
Formally, the ``states (\ref{5-17})" are specified by seven quantum numbers. 
However, the present system consists of four kinds of boson operators 
and the orthogonal set can be specified by four quantum numbers. 
Mainly, we discuss the meaning of this formal discrepancy. 
First, we investigate $({\hat Y}_{\pm,0})$ composed of 
$({\hat d}_0^* , {\hat d}_0)$. In this case, only one kind of 
boson operator is used, and then, the orthogonal set is given by 
\begin{equation}\label{6-1}
\drket{N}=({\hat d}_0^*)^N\drket{0} \ . \qquad (N=0,\ 1,\ 2, \cdots )
\end{equation}
On the other hand, the framework of the $su(1,1)$-algebra enable us to 
introduce the state 
\begin{equation}\label{6-2}
\drket{y,y_0}=({\hat Y}_+)^{y_0-y}\drket{y} \ . \qquad
(y_0=y+n,\ n=0,1,2.\cdots)
\end{equation}
Here, $\drket{y}$ is given in the relation (\ref{5-15}). 
Then, we have 
\begin{eqnarray}\label{6-3}
& &\drket{y=1/4,y_0}=({\hat Y}_+)^{n}\drket{0} \sim 
({\hat d}_0^*)^{2n}\drket{0} \ , \quad (N=2n) 
\nonumber\\
& &\drket{y=3/4,y_0}=({\hat Y}_+)^{n}{\hat d}_0^*\drket{0} \sim 
({\hat d}_0^*)^{2n+1}\drket{0} \ . \quad (N=2n+1) 
\end{eqnarray}
The eigenvalues of the Casimir operator ${\hat {\mib Y}}^2$ for the states 
$\drket{y=1/4,y_0}$ and\break
$\drket{y=3/4,y_0}$ are equal to the value $(-3/16)$. 
In the case $({\hat Z}_{\pm,0})$, the situation is the same as the 
case $({\hat Y}_{\pm,0})$. From the above argument, 
we learn that the quantum numbers $y$ and $z$ ($=1/4,\ 3/4)$ are 
used for discriminating between the even- and the odd-boson number states. 
Next, we investigate the case $({\hat W}_{\pm,0})$ composed of two kinds 
of bosons $({\hat c}_+ , {\hat c}_+)$ and $({\hat c}_-^*,{\hat c}_-)$. 
In this case, the orthogonal set is given by 
\begin{equation}\label{6-4}
\dkket{m_+,m_-}=({\hat c}_+^*)^{m_+}({\hat c}_-^*)^{m_-}
\dkket{0}\ . \quad 
(m,\ n=0,\ 1,\ 2,\ 3,\cdots)
\end{equation}
In the framework of the $su(1,1)$-algebra, we can introduce 
the following state : 
\begin{equation}\label{6-5}
\dkket{w(s),w_0}=({\hat W}_+)^{w_0-w}\dkket{w(s)} \ . \quad
(w_0=w+l,\ l=0,\ 1,\ 2,\cdots)
\end{equation}
Here, $\dkket{w(s)}$ is given in the relation (\ref{5-11}). 
Then, we have 
\begin{eqnarray}
& &\dkket{w(+),w_0}=({\hat c}_+^*)^{l+2w-1}({\hat c}_-^*)^l\dkket{0} \ , 
\quad (m_+=l+2w-1,\ m_-=l) 
\label{6-6}\\
& &\dkket{w(-),w_0}=({\hat c}_+^*)^{l}({\hat c}_-^*)^{l+2w-1}\dkket{0} \ , 
\quad (m_+=l,\ m_-=l+2w-1) 
\label{6-7}
\end{eqnarray}
The state (\ref{6-6}) and (\ref{6-7}) tell us that the set (\ref{6-4}) 
is classified in terms of $(m_+>m_-,\ m_+\leq m_-)$ and 
for this classification, $s$ is used. 
The eigenvalue of ${\hat {\mib W}}^2$ for the state (\ref{6-5}) is equal to 
$w(w-1)$ in two groups. 
From the above argument, we know the role of the quantum numbers $s$, $z$ and 
$y$. In the ``states (\ref{5-17})", $w$, $x$, $t$ and $t_0$ play the basic 
role, but, $s$, $z$ and $y$ restrict to the ranges of $w$, $x$ and $t$, 
which can be seen in various relations. 

In conclusion, we give some remarks. 
In this paper, we discussed three problems :\break
(i) The eigenvalue 
problem in an unconventional form, (ii) the coupling scheme 
of the addition of two $su(1,1)$-spins and (iii) a concrete example. 
These will serve boson realizations of the $so(4)$- and the 
$so(3,1)$-algebra developed in the next paper.

\acknowledgement

The main part of this paper was completed when the authors S.N., Y.T. and 
M.Y. stayed at Coimbra in September of 2004. 
They express their sincere thanks to Professor J. da Provid\^encia, 
co-author of this paper, for his kind invitation.

\appendix
\section{The procedure for determining $f_n(t,m)$}

First, we note that it may be enough to treat the function $f_n(t,m)$ 
in the form of a polynomial in the $n$-th degree for $m$. 
If it is accepted, the relation (\ref{2-26}) can be expressed in the form 
\begin{eqnarray}\label{a1}
& &n(3\alpha_n-\alpha_{n-1})m^n
\nonumber\\
& &
+\left[\left((3n-1)\beta_n-n\beta_{n-1}\right)-\left(\!
\frac{n(n-1)}{2}\alpha_n+2n(n-1+t)\alpha_{n-1}\!\right)\right]m^{n-1}
\nonumber\\
& & +\cdots +2n(\gamma_n-(n-1+t)\gamma_{n-a})=0 \ . 
\end{eqnarray}
Here, $f_n(t,m)$ is expressed as 
\begin{equation}\label{a2}
f_n(t,m)=\alpha_n m^n+\beta_n m^{n-1}+\cdots +\gamma_n \ . 
\end{equation}
Since we have the relation (\ref{a1}) for any value of $m$, the coefficient 
of any term $m^k$ $(k=n,n-1,\cdots ,0)$ should be vanished. 
Then, we obtain the recursion formulae for 
$(\alpha_n,\beta_n, \cdots ,\gamma_n)$

Under the above idea, we set up $f_n(t,m)$ in the form 
\begin{equation}\label{a3}
f_n(t,m)=\sum_{r=0}^{n}\frac{n!}{r!(n-r)!}g_n^{(n-r)}(t)\left(
\frac{m}{3}\right)^r \ . 
\end{equation}
Substituting the expansion (\ref{a3}) into the relation (\ref{2-25}) 
and following the above-mentioned procedure, we have the following formulae : 
\begin{eqnarray}
& &g_n^{(0)}(t)-g_{n-1}^{(0)}(t)=0 \ , 
\label{a4}\\
& &(3n-r)g_n^{(r)}(t)-3(n-r)g_{n-1}^{(r)}(t)\nonumber\\
& &\quad
+(n-r)\sum_{k=1}^{r}\frac{(-)^k}{3^k (1+k)}\cdot
\frac{r!}{k!(r-k)!}g_n^{(r-k)}(t)-2r(t+n-1)g_{n-1}^{(r-1)}(t)=0 \ . 
\nonumber\\
& &\qquad\qquad\qquad\qquad\qquad\qquad\qquad\qquad\qquad\qquad\qquad
 (r=1,2,\cdots,n)
\label{a5}
\end{eqnarray}
Since $g_0^{(0)}(t)=f_0(t,m)=1$, the relation (\ref{a4}) gives us 
\begin{equation}\label{a6}
g_n^{(0)}(t)=1 \ . 
\end{equation}
For $r=1$, the relation (\ref{a5}) leads us to 
\begin{eqnarray}\label{a7}
& &(3n-1)g_n^{(1)}(t)-3(n-1)g_{n-1}^{(1)}(t) 
-\frac{n-1}{6}g_n^{(0)}(t)-2(t+n-1)g_{n-1}^{(0)}(t) = 0 \ . \qquad\quad
\end{eqnarray}
The relations (\ref{a6}) and (\ref{a7}) determine $g_n^{(1)}(t)$ in 
the form 
\begin{equation}\label{a8}
g_n^{(1)}(t)=\frac{13}{30}n+\left(t-\frac{13}{30}\right) \ . 
\end{equation}
For $r=2$, we have 
\begin{eqnarray}\label{a9}
& &(3n-2)g_n^{(2)}(t)-3(n-2)g_{n-1}^{(2)}(t) \nonumber\\
& &\quad
+(n-2)\left(-\frac{1}{3}g_n^{(1)}(t)+\frac{1}{27}g_n^{(0)}(t)\right)
-4(t+n-1)g_{n-1}^{(1)}(t) = 0 \ . 
\end{eqnarray}
The solution of the relation (\ref{a9}) is given as 
\begin{equation}\label{a10}
g_n^{(2)}(t)=\frac{169}{900}n^2+\left(\frac{13}{15}t+\frac{181}{540}\right)n 
+\left(t^2-\frac{11}{15}t+\frac{1591}{1350}\right) \ . 
\end{equation}
In such a manner as the above, we obtain the form of 
$g_n^{(r)}(t)\ (r=3,4,\cdots, n-1)$. 
The case $r=n$, the relation (\ref{a5}) gives us 
\begin{equation}\label{a11}
2ng_n^{(n)}(t)-2n(t+n-1)g_{n-1}^{(n-1)}(t)=0 \ . 
\end{equation}
The relation (\ref{a11}) determines $g_n^{(n)}(t)$ in the form 
\begin{equation}\label{a12}
g_n^{(n)}(t)=\frac{\Gamma(t+n)}{\Gamma(t)} \ . 
\end{equation}
Concerning $g_n^{(r)}(t)$, the above examples suggest us the following form : 
\begin{eqnarray}
& &g_n^{(r)}(t)=\sum_{k=0}^{r}p_{n,r}^{(k)}(t)n^k \ , 
\label{a13}\\
& &p_{n,r}^{(k)}(t)=\sum_{l=0}^{k}q_{n,r,k}^{(l)}t^l \ . 
\label{a14}
\end{eqnarray}
The relations for obtaining $p_{n,r}^{(k)}(t)$, that is, $q_{n,r,k}^{(l))}$ 
are omitted.


\begin{thebibliography}{99}
\bibitem{1}
A. Klein and E. R. Marshalek, Rev. Mod. Phys. {\bf 63} (1991), 375. 
\bibitem{2}
S. Y. Li, A. Klein and R. M. Dreizler, J. Math. Phys. {\bf 11} (1970), 975.\\
F. Sakata, Y. Hashimoto, T. Marumori and T. Une, Prog. Theor. Phys. {\bf 70} 
(1981), 163.\\
M. Matsuo and K. Matsuyanagi, Prog. Theor. Phys. {\bf 74} (1985), 288.\\
K. Takada, Y. R. Shimizu and H. Thorn, Nucl. Phys. {\bf A 485} (1988), 187.\\
M. Sambataro, Phys. Rev. {\bf C 60} (1999), 0643200.
\bibitem{3}
M. O. Terra, M. C. Nemes, C. Provid\^encia and J. da Provid\^encia, 
Ann. of Phys. {\bf 262} (1998), 1. 
\bibitem{4}
E. Celeghini, M. Rasetti, M. Tarlini and G. Vittiello, Mod. Phys. Lett. 
{\bf B 3} (1989), 1213.\\
E. Celeghini, M. Rasetti and G. Vittiello, Ann. of Phys. {\bf 215} (1992), 156.
\bibitem{5}
Y. Tsue, A. Kuriyama and M. Yamamura, Prog. Theor. Phys. {\bf 91} (1994), 
49 ; 469.\\
A. Kuriyama, J. da Provid\^encia, Y. Tsue and M. Yamamura, Prog. Theor. Phys. 
Suppl. No. 141 (2001), 113. 
\bibitem{6}
A. Kuriyama, J. da Provid\^encia and M. Yamamura, Prog. Theor. Phys. {\bf 103} 
(2000), 285; 305.\\
M. Yamamura, A. Kuriyama and T. Kunihiro, Prog. Theor. Phys. {\bf 104} 
(2000), 385 ; 401. \\
A. Kuriyama, C. Provid\^encia, J. da Provid\^encia, Y. Tsue and M. Yamamura, 
Prog. Theor. Phys. {\bf 109} (2003), 959. 
\bibitem{7}
A. Kuriyama, C. Provid\^encia, J. da Provid\^encia and M. Yamamura, 
Prog. Theor. Phys. {\bf 104} (2000), 143. 
\end{thebibliography}
\end{document}